\documentclass[preprint,superscriptaddress,amsmath,amssymb,aps,longbibliography,nofootinbib]{revtex4-1}

\usepackage{comment,xcolor}

\usepackage{bbold}
\usepackage{graphicx}
\usepackage{dcolumn}
\usepackage{bm}
\usepackage{setspace}
\usepackage[caption=false]{subfig}
\usepackage{mathrsfs}
\usepackage{hyperref}

\newcommand{\be}{\begin{equation}}
\newcommand{\ee}{\end{equation}} 

\AtBeginDocument{%
    \newwrite\bibnotes
    \def\bibnotesext{Notes.bib}
    \immediate\openout\bibnotes=\jobname\bibnotesext
    \immediate\write\bibnotes{@CONTROL{REVTEX41Control}}
    \immediate\write\bibnotes{@CONTROL{%
    apsrev41Control,author="08",editor="1",pages="1",title="0",year="1"}}
     \if@filesw
     \immediate\write\@auxout{\string\citation{apsrev41Control}}%
    \fi
}%

\begin{document}

\title{Decoherence from Long-Range Forces in Atom Interferometry}

\author{Jonathan Kunjummen}
\email{jkunjumm@umd.edu}
\affiliation{%
Joint Quantum Institute/Joint Center for Quantum Information and Computer Science,
University of Maryland, College Park/National Institute of Standards and Technology, Gaithersburg, MD, USA}%
\author{Daniel Carney}%
\affiliation{%
Physics Division, Lawrence Berkeley National Laboratory, Berkeley, CA, USA}%
\author{Jacob M. Taylor}
\affiliation{%
Joint Quantum Institute/Joint Center for Quantum Information and Computer Science, University of Maryland, College Park/National Institute of Standards and Technology, Gaithersburg, MD, USA}%

\date{\today}
  
\begin{abstract}
Atom interferometers provide a powerful means of realizing quantum coherent systems with increasingly macroscopic extent in space and time. These systems provide an opportunity for a variety of novel tests of fundamental physics, including ultralight dark matter searches and tests of modifications of gravity, using long drop times and microgravity environments. However, as experiments operate with longer periods of free fall and become sensitive to smaller background effects, key questions start to emerge about the fundamental limits to future atom interferometery experiments. We study the effects on atomic coherence from hard-to-screen backgrounds due to baths of ambient particles with long-range forces, such as gravitating baths and charged cosmic rays. Our approach -- working in the Heisenberg picture for the atomic motion -- makes proper inclusion of the experimental apparatus feasible and clearly shows how to handle long-range forces. We find that these potential backgrounds are likely negligible for the next generation of interferometers, as aggressive estimates for the gravitational decoherence from a background bath of dark matter particles gives a decoherence timescale on the order of years.
\end{abstract}

\maketitle

\begin{spacing}{1}
\tableofcontents
\end{spacing}

\section{Introduction}\label{sec:Intro}
Atom interferometers (AIs) are useful both in enabling new measurements and in providing better understandings of key concepts in quantum mechanics and its intersection with relativity~\cite{Cronin2009}. Based on the concept of particle interference, the simplicity of the atom interferometer in principle provides key ways of testing the connection between path integrals and Hamiltonian quantum dynamics~\cite{Storey1994}, explorations of the concept of `which-way' information~\cite{Scully1991,Bjork1998,Durr1998,Durr2000,Hornberger2003b,Marzlin_2008}, and considerations of the distinctions between interferometry in a Galilean frame and in a fully general relativsitic frame~\cite{Dimopoulos2008a}. At the same time, AIs are exquisitely sensitive to small variations in local accelerations and rates of rotation. Early work showcased measurements of the tides~\cite{Peters2001} and seismic backgrounds~\cite{Hu2013,Freier2016} while more recent developments include precision geodesy and gravity gradient detection~\cite{McGuirk2002,Stockton2011,Asenbaum2017}. In the next decade, AIs will see increasing use in searches for new fundamental physics, including low-frequency gravitational waves and ultralight dark matter \cite{Hogan2011,Graham:2015ifn,canuel2018exploring,Badurina2020,Abe2021} and tests of modifications of gravity \cite{hamilton2015atom,burrage2015probing}.

Typically, the dominant background noise in AI systems is assumed to be a combination of different terrestrial sources, due to local gas particles in the (imperfect) vacuum, or due to laser-related noise sources~\cite{peters2001high}. Here, we instead consider the effect of both distant and close gravitational sources that may be due, e.g., to particulate models of dark matter or more distant, heavier astrophysical objects \cite{Fedderke2020}. Due to the unscreened, long-range nature of the gravitational coupling, this type of background is unavoidable for AIs. By examining this potential signal, we can put bounds on the fundamental performance limits of AIs in both earth-based and space-based settings, conditional on a variety of models for background particles.  This work also contributes to the study of irreducible gravitational backgrounds that optomechanical experiments probing quantum macroscopicity and gravity from quantum objects will face \cite{Carney2019,Carney2021,Bose2017,Toros2020,Gerlich2011,Rodewald2018,Delic2020,Westphal2021,Oniga2016,Pang2016,Pikovski2015,Bassi2017,Anastopoulos2013,Wang2006,Reynaud2001,Kok2003,Lamine2002}. 

In addition to practical reasons for wanting to characterize this background, the case of gravitational noise is conceptually interesting, firstly because of equivalence principle considerations. Typically, to account for background particles one need only include their interaction with the atom \cite{Gallis1990,Hornberger2003a,Adler2006}. Things become more complicated when we consider gravitational interactions, however, since for instance we know that a constant gravitational field acts on all components of an experiment so as to be undetectable in freefall. To capture this in the formalism describing the experiment, one must account for the effect of the gravitational field on the control system, not just the atom \cite{Dimopoulos2008a,jaekel2013phases,Toros2020}. A central result of our work is to provide a clean conceptual framework that treats the AI apparatus and the atoms themselves on an equal footing, thus properly treating issues involving the equivalence principle, which is an essential first step for taking AI into the space-based regime. The effect of gravity gradients, particularly seismic noise, on interferometry has also been analyzed by the gravitational wave community \cite{hughes_seismic_1998,thorne_human_1999,Pitkin_2011,abbott_ligo_2009}, motivating the construction of gravitational wave detectors in space \cite{armano_sub-femto-_2016,amaro-seoane_laser_2017}. However, while the LISA Pathfinder mission measured favorable acceleration noise rates for gravitational wave detection \cite{armano_sub-femto-_2016}, it attributed most of the noise to non-gravitational sources, and as a result some of the measured background depends on the nature of the test masses and the experimentalist's ability to screen unwanted interactions. Here, we prove a framework for calculating the inherent gravitational background any such experiment faces, showing how the ultimate noise level depends on characteristics of the bath.

The second piece of conceptual interest for this work comes from the infinite scattering cross section of the $1/r$ potential. Our approach addresses a conceptual issue which occurs when trying to apply traditional Brownian motion calculations \cite{Gallis1990,Hornberger2003a,Adler2006} when the bath has a long-range, unscreened coupling to the apparatus. As we will explain, the naive prediction of an infinite decoherence rate for a $1/r$ potential is fixed by including the infrared cutoff set by the spatial extent of the experimental apparatus \cite{Weinberg:1965nx,Carney:2017jut}. As we show below, the effect is likely to be small for models of dark matter under current consideration, but this study opens the door to exploring dark matter or other particulate detection using AIs when the interaction is not (only) gravitational, and to this end we consider interactions with cosmic ray particles at the end of the paper.

The paper is organized as follows. In Section \ref{sec:Bkgd} we set up the model for an atom interferometer in the presence of background particles that interact with a long-range force with both the atoms and the experimental apparatus. In Section \ref{sec:Experiment} we set up the formalism describing our atom interferometry experiment, and show how the effects of a constant, global acceleration on the atom and on the control system cancel out, in accordance with the Equivalence Principle.  This allows us to we analyze the effect of a cloud of distant bath particles on the experiment. In Section \ref{sec:CD}, we discuss the relationship between this calculation and the standard quantum Brownian motion framework. In Section \ref{sec:Near} we calculate the effect of nearby bath particles, in particular considering atom interferometers as dark matter detectors. In Section \ref{sec:Tuning} we discuss the usefulness of tuning the atom-laser distance to change sensitivity of the experiment to the gravitational background. In Section \ref{sec:Cosmic} we consider atom interferometers as impulse detectors for more general forces, examining the case of a passing charged cosmic ray particle. Finally, we summarize our results and consider future work in Section \ref{sec:Conclusion}.

\begin{figure}

\subfloat[]{
    \includegraphics[width=.3\linewidth]{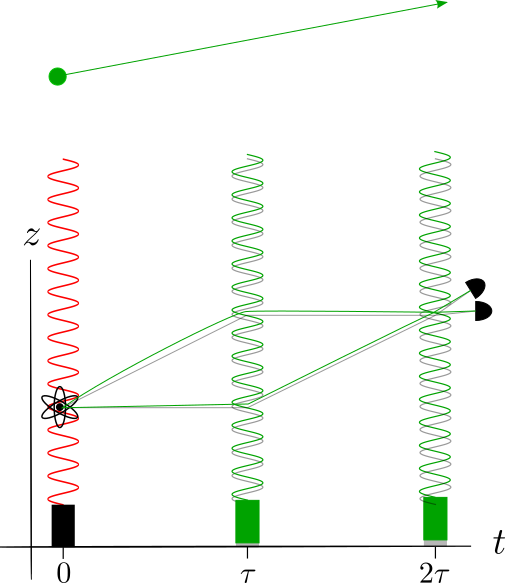}\label{fig:Fig1b}
}
\hspace{5pt}
\subfloat[]{
    \includegraphics[width=.3\linewidth]{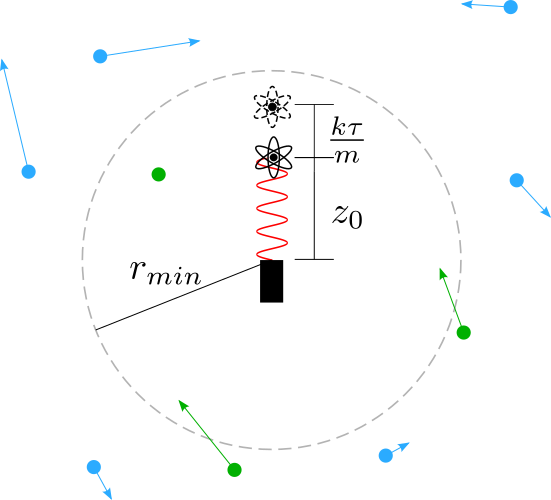}\label{fig:Fig1a}
}
\hspace{5pt}
\subfloat[]{
    \includegraphics[width=.3\linewidth]{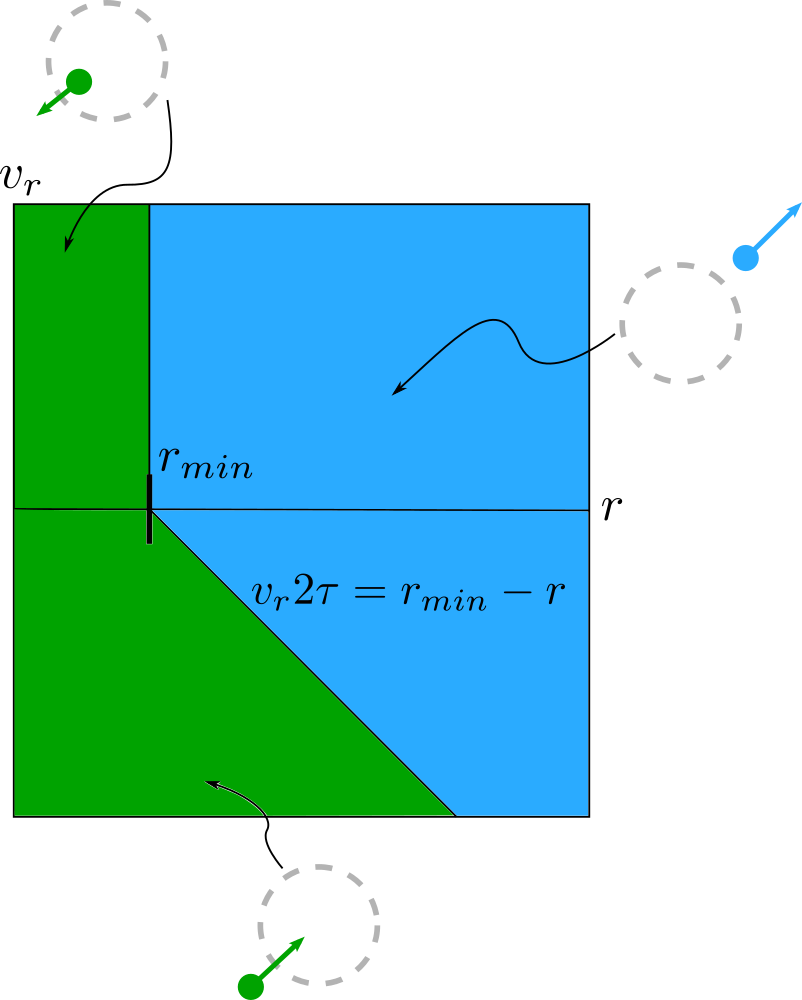}\label{fig:Fig1c}
}
\caption{Schematic of atom interferometry taking place in the presence of background particles. a) Schematic of atom interferometry. The atom begins in the ground state and then undergoes a $\frac{\pi}{2} - \pi - \frac{\pi}{2}$ pulse sequence, after which the internal state is measured. A passing bath particle leads to distortions of the laser and atom trajectories. b) Particles which remain outside $r_{\rm min}$ through the whole experiment (light blue) source a gravitational potential varying slowly across the spatial extent of the experiment. Those which start or drift inside $r_{\rm min}$ are in the collision cone (green). The atom starts a distance $z_0$ from the laser, and is put into a superposition of paths with maximum separation $\hbar k\tau/m$. c) Division of phase space into distant sector (blue) and collision cone (green). This is a cross section of the full phase space at purely radial velocity and arbitrary angular coordinates. The collision cone includes particles which begin inside the cutoff radius $r_{\rm min}$, as well as those which begin outside but have an inward velocity large enough to bring them within $r_{\rm min}$ by time $2\tau$.}
\end{figure}

\begin{figure}
\subfloat[]{
    \includegraphics[width=.3\linewidth]{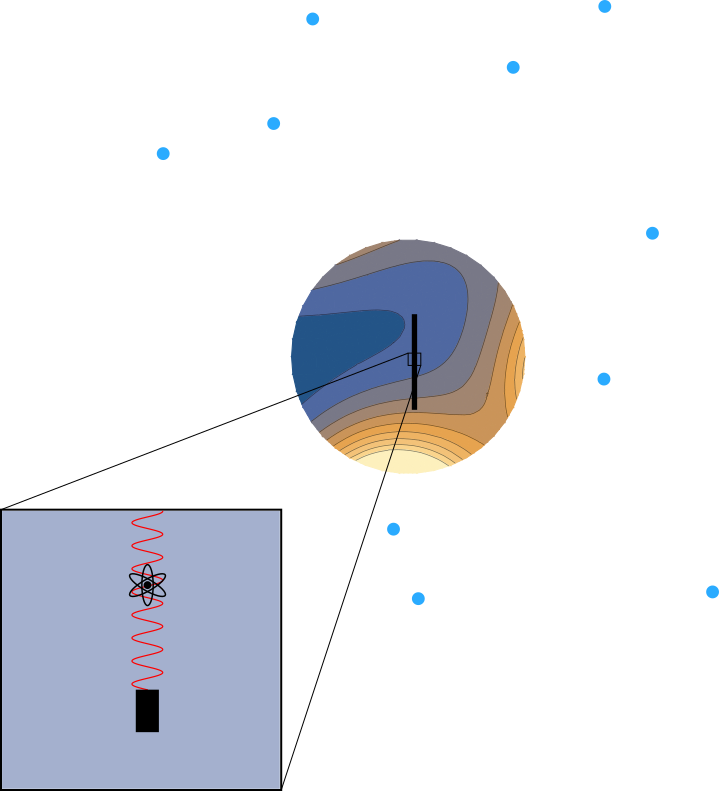}\label{fig:Fig2a}
}
\hspace{.5cm}
\subfloat[]{    
    \includegraphics[width=.45\linewidth]{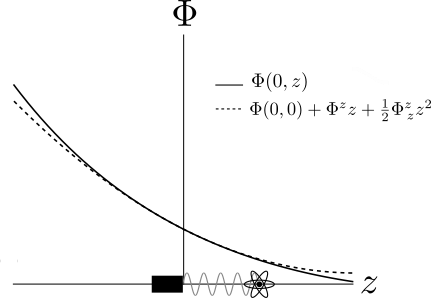}\label{fig:Fig2b}
}
\caption{a) Contour plot in two dimensions of the gravitational potential sourced by particles outside the cutoff radius $r_{\rm min}$. b) Comparison of the actual potential and the second order multipole approximation to it near the origin along the one dimensional slice shown in part a).}
\end{figure}
\section{Background Model}\label{sec:Bkgd}
In this paper we consider a simplified atom interferometer in microgravity, as shown in Fig. \ref{fig:Fig1b}, where the dominant gravitational acceleration is from a cloud of $N$ identical background particles of mass $m_b$ interacting with the setup via gravity. This model is relevant to the behavior of proposed atom interferometers in space, where the free evolution times are not limited by the need to construct a long drop tower \cite{Hogan2011}.  We restrict ourselves to a highly simplified atom interferometry model, in which atomic motion takes place along only one axis and there is only a single control laser, in order to focus on the effect of gravitational noise. Specifically, as described below, this interferometer is sensitive to local accelerations. A uniform acceleration introduces no relative motion between the control apparatus and the atom, and is therefore unobservable. Gravitational noise results, then, when an unknown background gravitational field introduces relative motion between any of three parts of the experiment: the control system and either of the paths taken by the atom in superposition. To model this carefully, and in contrast to prior work, we explicitly include in our analysis the motion of the laser which implements the beamsplitter and mirror operations.
\newline \indent
For simplicity, we ignore interactions between the bath particles, and assume that at the beginning of the experiment they are statistically distributed according to a free particle Boltzmann distribution with some effective temperature $T$
\begin{equation}
    P(\mathbf{r_b}, \mathbf{v_b}) = \frac{1}{V} \Big(\frac{m_b}{2\pi \,k_B T}\Big)^{-\frac{3}{2}} \exp\left\{-\frac{m_b \mathbf{v_b}^2}{2k_B T}\right\}
\end{equation}
where $k_B$ is Boltzmann's constant, and $V$ is the volume of all space. We will eventually take the large $N$ and large $V$ limit, with finite number density $n_0 \equiv N/V$.
\newline\indent
If a particle is much farther from the setup than the atom-laser distance, then the potential it produces near the experiment can be approximated with a multipole expansion whose first order terms constitute a global acceleration. We therefore divide the phase space of a bath particle into two sectors by choosing some cutoff radius $r_{\rm min}$ much larger than the spatial extent of the experiment, and then separating the part of phase space which will stay outside this radius for the entire duration of the experiment from the part which is inside $r_{\rm min}$ during some or all of the experiment. This division, which depends on the experimental runtime, is illustrated graphically in Fig. 1c. We will examine the effect of each of these phase sectors on an atom interferometry experiment in turn. In order to calculate what the effect of this is on atom interferometry, we need to define the experimental procedure in detail.

\section{Prototypical Experiment}\label{sec:Experiment}

We analyze a pedagogical model of atom interferometry to focus on the effects of gravitational noise. Specifically, we consider an experiment with two components, a laser of mass $M_{\rm laser}$ and position $\mathbf R_{\rm laser}$ and a two-level atom of mass $m_a$ and position $\hat{\mathbf r}$. 
The laser produces light with wavevector $\mathbf{k} = k \,\mathbf{e_3}$ which we treat classically. We emphasize that for the purposes of this calculation, we neglect the spatial distribution and relative motion of different components of the laser system. Assuming that all effects leading to phase shifts in the laser light are captured by a single gravitational potential acting on the laser center of mass, which we write as $-G M_{\rm laser} \Phi(\mathbf{R}_{\rm laser})$ with $G$ the gravitational constant, the evolution of the laser is governed by the classical equations of motion
\begin{align}
    \dot{\mathbf{R}}_{\rm laser} &= \mathbf{P}_{\rm laser}/M_{\rm laser}\label{eqn:CEOM1}\\
     \dot{\mathbf{P}}_{\rm laser}&= G M_{\rm laser} \nabla \Phi(\mathbf{R}_{\rm laser})\label{eqn:CEOM2}.
\end{align}
The evolution of the atom degrees of freedom is governed by the following Hamiltonian, making the rotating wave approximation and moving to the interaction picture with respect to the atom's internal degree of freedom:
\begin{align}
&\hat H = \hat H_{\rm atom} + \hat H_{\rm int}\\
&\hat H_{\rm atom} =   \frac{\hat{\mathbf{p}}^2}{2m_a} -G m_a \hat \Phi(\hat {\mathbf{r}})- \frac{ \hbar \Delta}{2}\hat \sigma_z\\
&\hat H_{\rm int} = \frac{\hbar \Omega(t)}{2} \,e^{i k \big(\hat z-Z_{\rm laser}(t)\big)-i\Delta t}\, \hat\sigma_+ + \text{h.c.},\qquad
\hat \sigma_+ = \hat \sigma_-^\dag=|\!\uparrow\rangle \langle\downarrow\!|,
\end{align}
\noindent where $\Omega (t)$ is the slowly varying Rabi frequency, $\sigma_+$ ($\sigma_-$) is the atomic raising (lowering) operator, and h.c. stands for Hermitian conjugate. We stress that the gravitational potential which we, after factoring out the gravitational constant and the mass, denote $\Phi$ has the same functional form for both the atom and the laser, which will lead to the noise cancellations associated with the equivalence principle. From now on we work on resonance, setting $\Delta =0$ in what follows. 
\newline \indent 
Notice we make the approximation that the magnitude of the coupling between the atom and the laser field depends only on time, not on the location of the atom in space. We also ignore time retardation effects in the location of the laser seen by the atom. These approximations are valid if the laser beam is much broader than the atom cloud at all times, and if the time it takes light to travel from the laser to the atom is much faster than the timescale on which the laser amplitude changes.
\newline \indent
To implement our experimental protocol, for simplicity we take the pulse envelope $\Omega(t)$ to be a sequence of delta functions giving a three pulse sequence 
\begin{equation}
    \Omega(t) = \frac{\pi}{2} \delta(t)+e^{-i\theta_0/2}\,\pi\, \delta(t-\tau) +\frac{\pi}{2} \delta(t-2\tau) , \label{eqn:pulses}
\end{equation}

\noindent where $\theta_0$ is a tunable phase set by the experimenter and $\tau$ is the time between pulses. We can approximate the pulses as delta functions as long as the pulses are much faster than the dynamics of the laser and atom position. After the pulse at time $2\tau$ we then measure the state of the atom in the $\uparrow,\downarrow$ basis.
\newline\indent
We focus on making measurements of the internal state of the atom, using a spinor notation so that $\big|\chi \big\rangle$ denotes the full spinor state, whereas $|\psi\rangle$ denotes a state in real space only. In this way we write
\begin{align}
    \big|\chi\big\rangle = 
    |\psi_\uparrow\rangle\otimes|\!\uparrow\rangle +
    |\psi_\downarrow\rangle\otimes|\!\downarrow\rangle
    =
    \begin{pmatrix}
    |\psi_\uparrow\rangle \\
    |\psi_\downarrow\rangle
    \end{pmatrix}
\end{align}
with normalization condition,
\begin{align}
    1 = \big\langle\chi\big|\chi \big\rangle = \langle \psi_\uparrow|\psi_\uparrow \rangle +\langle \psi_\downarrow|\psi_\downarrow\rangle =\int d^3 \mathbf{r}\Big( |\psi_\uparrow(\mathbf{r})|^2+|\psi_\downarrow(\mathbf{r})|^2\Big)
\end{align}
so that $|\psi_\uparrow(\mathbf{r})|^2$ gives the probability density for finding the atom in the excited state at $\mathbf r$ and likewise $|\psi_\downarrow (\mathbf r)|^2$ for the ground state. Generally, we will assume measurements do not resolve atomic position and thus we only have access to the probabilities after integration
\begin{equation}
    \text{Prob}(\uparrow) = \int d^3 \mathbf{r} \,|\psi_\uparrow (\mathbf r)|^2
\end{equation}
and likewise for spin down.
\newline \indent 
To calculate the experimental signal, we work in terms of the unitary operator generated by the Hamiltonian. This approach is complementary to the path integral method, used widely to describe atom interferometers \cite{refId0}. Our current approach is convenient, however, in that it easily incorporates laser motion, failure of the two paths to interfere, and (as will be shown in Section \ref{sec:Near}) impulses from nearby bath particles which perturb the atomic trajectories significantly. Given the pulse sequence of Eqn. \ref{eqn:pulses}, the solution to the Schrodinger Equation can be expressed in terms of a few unitary operators, first the unitary corresponding to free evolution between pulses
\be
U_{\rm atom}(t_1,t_2) = \mathcal{T}\exp\left\{-i\int_{t_1}^{t_2} \frac{dt H_{\rm atom}(t)}{\hbar}\right\}
\ee
and next the unitaries corresponding to the instantaneous $\pi/2$ and $\pi$ pulses (we neglect the effect of gravity during the pulses)
\begin{equation} \label{eqn:piby2pulse}
    U_{\pi/2}(t) \equiv \exp\left\{-i\Big(\frac{\pi}{4}e^{ik(\hat z-Z_{\rm laser}(t))} \sigma_++\text{h.c.}\Big)\right\} = \frac{1}{\sqrt{2}}\mathbb{1}-\frac{i}{\sqrt{2}}\big(e^{i k(\hat z-Z_{\rm laser}(t))}\sigma_+ + \text{h.c.} \big)
\end{equation}
\begin{equation} \label{eqn:pipulse}
    U_{\pi}(t) \equiv \exp\left\{-i\Big(\frac{\pi}{2}e^{i(k(\hat z-Z_{\rm laser}(t))-\theta_0/2)} \sigma_++\text{h.c.}\Big)\right\} = -\frac{i}{\sqrt{2}}\big(e^{i (k(\hat z-Z_{\rm laser}(t))-\theta_0/2)}\sigma_+ + \text{h.c.} \big)
\end{equation}
where $Z_{\rm laser}(t)$ is of course evolved by the classical equations of motion, Eqns. \ref{eqn:CEOM1}, \ref{eqn:CEOM2}.
\newline \indent 
The spinor state at the end of the protocol is
\begin{align}
\big|\chi(2\tau)\big\rangle = U_{\pi/2}(2\tau)\,U_{\rm atom}(\tau,2\tau)\,U_{\pi}(\tau)\,U_{\rm atom}(0,\tau)\,U_{\pi/2}(0)\big|\chi(0)\big\rangle
\end{align}
So, taking the initial atomic state to be entirely in the ground state with spatial wavefunction $|\psi_{\downarrow,t=0}\rangle$
\be
\big|\chi(0)\big\rangle = 
\begin{pmatrix}
0\\
|\psi_{\downarrow,\,t=0}\rangle
\end{pmatrix}
\ee
we get the final spinor state
\begin{equation}\label{eqn:spinorFinal}
    \big|\chi(2\tau)\big\rangle =    \frac{1}{2}
    \begin{pmatrix}
    ie^{i k (\hat z-Z_{\rm laser}(2\tau))}\,(U^t-U^b)\;|\psi_{\downarrow,\,t=0}\rangle\\
    -(U^t+U^b)\;|\psi_{\downarrow,\,t=0}\rangle
    \end{pmatrix},
\end{equation}
where we have defined the operators
\begin{align}
U^t &= e^{i\frac{\theta_0}{2}}\,U_{\rm atom}(\tau,2\tau)\,e^{-i k(\hat z -Z_{\rm laser}(\tau))}\,U_{\rm atom}(0,\tau)\,e^{i k \hat z} \label{eqn:Utop}\\
U^b &= e^{-i\frac{\theta_0}{2}}\,e^{-i k (\hat z - Z_{\rm laser}(2\tau))}\,U_{\rm atom}(\tau,2\tau)\,e^{i k (\hat z  - Z_{\rm laser}(\tau))}\,U_{\rm atom}(0,\tau) \label{eqn:Ubot}
\end{align}
which act only on the center of mass degree of freedom. 
\newline\indent 
The two unitaries $U^t, U^b$ correspond to the two arms of the interferometer as shown in Fig. \ref{fig:Fig1b} \cite{Schleich2013}. To provide some intuition, there are two ways an atom can end up in the ground state. The ground state contribution from the top arm (described by $U^t$) is excited by the first $\pi/2$ pulse, de-excited by the $\pi$ pulse, and left unchanged by the second $\pi/2$ pulse. The contribution from the bottom arm (described by $U^b$) corresponds to the component unchanged by the first $\pi/2$ pulse, excited by the $\pi$ pulse, and de-excited by the last $\pi/2$ pulse. These two contributions follow different real space paths, as demonstrated in the figure. These same operators $U^t, U^b$ are summed to calculate the final excited state wavefunction in the top entry of Eqn. \ref{eqn:spinorFinal}, but they are multiplied by a final momentum kick\textemdash as the excited state still carries the photon momentum\textemdash and summed with a different relative phase as is required to conserve probability. 
\newline
\indent
We can calculate the diagonal entries of the spin density matrix, i.e. the populations of the ground and excited state, at the end of the experiment
\begin{align}
    \rho_{\downarrow\downarrow} =1-\rho_{\uparrow \uparrow}&=\frac{1}{2}+\frac{1}{2} \,\text{Re}\,  \langle \psi_{\downarrow,\,t=0}|\, (U^b)^\dag U^t \,|\psi_{\downarrow,\,t=0}\rangle \label{eqn:Population}
\end{align}
and as expected, if the spatial wavefunctions at the end of the two arms do not overlap, i.e $\langle (U^{b})^\dag U^t\rangle = 0$, the contributions from the two arms cannot interfere, and both states will be equally populated. If there is no external potential, then $(U^b)^\dag U^t = e^{i\theta_0}$ and the output population is simply
\begin{align}\label{eqn:noiseless}
    \rho^{\rm free}_{\downarrow \downarrow} = \frac{1}{2}+\frac{1}{2}\cos\theta_0.
\end{align}
In this case, by tuning $\theta_0$, we can tune the final population all the way from entirely ground state to entirely excited state.
\newline \indent 
Gravitational noise changes the overlap between the two paths, and can cause both phase shifts in the cosine term and an overall reduction in contrast. We will quantify decoherence from gravitational noise by how it decreases our ability to tune the final population via $\theta_0$. We rearrange the terms in the path-specific unitaries to get
\begin{align}
U^t &= e^{i\frac{\theta_0}{2}}\,U_{\rm atom}(0,2\tau)\,e^{-i k(\hat z(\tau) -Z_{\rm laser}(\tau))}\,e^{i k \hat z} \\
U^b &= e^{-i\frac{\theta_0}{2}}\,U_{\rm atom}(0,2\tau)\,e^{-i k (\hat z(2\tau) - Z_{\rm laser}(2\tau))}\,e^{i k (\hat z(\tau)  - Z_{\rm laser}(\tau))} \end{align}
written in terms of the interaction picture position operator
\begin{equation}
    \hat z \to \hat z(t) = \Big(U_{\rm atom}(0,t)\Big)^\dag \hat z \,U_{\rm atom}(0,t)
\end{equation}
we can then rewrite the overlap factor in the interaction picture
\begin{align}
    \Big(U^b\Big)^\dag U^t&= \Big(U_{\rm atom}^\dag(0,2\tau)\,U^b\Big)^\dag U_{\rm atom}^\dag(0,2\tau) U^t\nonumber\\[.1cm]
    &=e^{i\theta_0} \Big(e^{-i k (\hat z(2\tau) - Z_{\rm laser}(2\tau))}\,e^{i k (\hat z(\tau)  - Z_{\rm laser}(\tau))}\Big)^\dag \,e^{-i k(\hat z(\tau) -Z_{\rm laser}(\tau))}\,e^{i k \hat z} \label{eqn:intOverlap}
\end{align}
From this we see that the output of the atom interferometer depends solely on the evolution of the position operator $\hat z(t)$ relative to the laser position. Thus, shared acceleration is unobservable in this setup, as expected from the equivalence principle.

\subsection{Case Study: Uniform Acceleration}
To illustrate the importance of including the effect of gravity on the control system, before going further with our model of the gravitational noise background we consider an experiment subject to global acceleration $g$, i.e.
\begin{align}
    G\,\Phi (\mathbf R_{\rm laser})= g \,Z_{\rm laser}
\end{align}
and an analogous expression holds for the atom with the appropriate quantities promoted to operators. The classical and Heisenberg equations of motion in this case can be solved exactly to give the evolution of the laser position and atom center of mass
\begin{align}
    &Z_{\rm laser}(t) = Z_{\rm laser}(0)+\dot Z_{\rm laser}(0) t +\frac{1}{2}gt^2\\
    &\hat{z}(t) = \hat z + \hat p_z t/m_a +\frac{1}{2}gt^2.
\end{align}
We will always choose our coordinate system so that the laser starts out at rest at the origin, $Z_{\rm laser}(0),\,\dot Z_{\rm laser}(0)=0$. We subsequently can evaluate Eqn. \ref{eqn:intOverlap} with
\begin{align}
    \hat{z}(t) - Z_{\rm laser}(t) = \hat z + \hat p_z t/m_a.
\end{align}
But of course, this is how the relative displacement would evolve if the apparatus was floating in free space. 
The output spin down population is again given by Eqn. \ref{eqn:noiseless} with no dependence on the global acceleration $g$, so indeed the equivalence principle is satisfied. We now have the tools in place to explore our gravitational noise model. We first examine the distant sector.

\subsection{Distant Sector}\label{sec:Distant}
We now analyze the bath model introduced earlier. As the $n$th bath particle at position $\mathbf r_{b_n}$ produces a potential given by the Newton formula, the total potential is simply the sum over individual contributions
\begin{align}
G \hat{\Phi}(\hat{\mathbf r}) &\equiv G\,m_b\sum_{n=1}^N \frac{1}{|\mathbf r_{b_n}-\hat{\mathbf r}|}.
\end{align}
Bath particles in the distant sector produce a gravitational potential near the experiment that can be approximated with a multipole expansion as shown in Figs. \ref{fig:Fig2a}-\ref{fig:Fig2b},
\begin{align}
    \hat{\Phi}(\hat{\mathbf r}) &\approx \Phi(\mathbf 0) +\Phi^i \hat r_i+\frac{1}{2}\hat r_i \,\Phi^i_{\,\,j} \,\hat r^j \label{eqn:Multipole} 
\end{align}
where repeated indices are summed over the three spatial axes, but as this is a nonrelativistic calculation there is no distinction between covariant and contravariant vectors. We have kept the terms leading to a global acceleration
\begin{equation}
    \Phi^i \equiv  \frac{\partial}{\partial \hat r_i} \hat \Phi(\hat{\mathbf{r}})\Big|_{\hat {\mathbf{r}}=\mathbf 0}   
\end{equation}
and the quadratic terms 
 \begin{equation}
     \Phi^{i}_{\;j} \equiv \frac{\partial^2}{\partial \hat r_i\partial \hat r_j} \hat \Phi(\hat {\mathbf{r}})\Big|_{\hat{ \mathbf{r}}=\mathbf 0}.
 \end{equation}
We emphasize, too, that although $ \hat{\Phi}(\hat{\mathbf r})$ is an operator, the coefficients $\Phi^i, \,\Phi^i_{\,\,j}$ are real numbers. 
\newline \indent 
We can calculate the explicit form of the quadratic coefficients $\Phi^{i}_{\;j}$
\begin{align}
    \Phi^z_{\;z} \equiv \,m_b\sum_{n=1}^N \frac{-r_{b_n}^2 + 3 z_{b_n}^2}{r_{b_n}^5}\\
    \Phi^z_{\;y} \equiv \,m_b\sum_{n=1}^N \frac{3 z_{b_n} y_{b_n}}{r_{b_n}^5},
\end{align}
with rotational symmetry fixing all other components given these two. Following the argument in the uniform acceleration case study, we note that the $\Phi^i$ terms are not observable as they produce identical displacement of the laser and the atom, therefore we do not bother to list them explicitly. Because we assume the different bath particles are distributed independently, any function $f(\Phi^{i}_{\;j})$ has expectation value
\begin{align}
    \langle f(\Phi^{i}_{\;j}) \rangle =  \frac{1}{V^N} \left(\frac{m_b}{2\pi k_B T}\right)^{\frac{3N}{2}}\iint\limits_{\text{distant}} d^{3N} \{\mathbf{r}_{b_n}\}\,  d^{3N} \{\mathbf{v}_{b_n}\}     \exp\{-\frac{m_b}{2 k_B T}\sum_n\mathbf{v}_{b_n}^2\} f(\Phi^{i}_{\;j}(\{\mathbf{r}_{b_n}\}, \{\mathbf{v}_{b_n}\}, t)),
\end{align}
where the subscript on the integrals indicates that we are averaging over the distant phase space sector as shown in Fig. 1c.
\newline\indent 
In general the $\Phi^i_{\,j}$ are time dependent because of the motion of the bath particles, but in this section we will assume the zero temperature limit for simplicity. We calculate the lowest order correction from bath particle motion in Appendix \ref{app:timeDep}. In the zero temperature case, all background particles are at rest, and the collision cone particles are simply those that begin inside the cutoff radius. For any finite thermal bath velocity $v_\beta$, the zero temperature approximation gives increasingly accurate behavior for the distant sector (though not for the collision cone) as we take $r_{\rm min} \to\infty$, because changes in the gravitational field near the origin over the course of the experiment are suppressed by $v_\beta \tau/ r_{\rm min}$. We can drop the velocity averaging in the zero temperature limit and simply average over possible bath particle locations. If the average volume per bath particle is much smaller than the excluded volume, $n_0 r_{\rm min}^3 \gg 1 $, the components of $\Phi^{i}_{\;j}$ are approximately Gaussian distributed with zero mean and variance
\begin{align}\label{eqn:curvVariance}
    \langle (G\, \Phi^{i}_{\;j})^2 \rangle = \frac{4\pi}{3}\frac{3+\delta_{ij}}{5}\xi^2,\qquad 
    \xi^2 \equiv \frac{(Gm_b)^2n_0}{r_{\rm min}^3}.
\end{align}
We point out that the characteristic fluctuation scale  $\xi^2$ has dimension $[1/\text{time}^{4}]$. The trace of the matrix of quadratic terms is identically zero, so some of its components are correlated. We will skip over this subtlety, however, as only an uncorrelated set of components ends up contributing in the lowest order correction to the flat space behavior.
\newline \indent 
As in the global acceleration case, we solve the Heisenberg and classical equations of motion for $\hat z, Z_{\rm laser}$ respectively to calculate the evolution of the laser-atom separation 
\begin{align}
    \hat z(t) - Z_{\rm laser}(t) \approx  \hat z + t\,\frac{\hat p_z}{m_a} +   \frac{t^2}{2!}G\,\Phi^{z}_{\;j}\,\hat r^j +\frac{t^3}{3!}G\,\Phi^{z}_{\;j}\,\frac{\hat p^j}{m_a} +\frac{t^4}{4!}G^2\,\Phi^{z}_{\;j}\,\Phi^{j}_{\;k}\,\hat r^k+\frac{t^5}{5!}G^2\,\Phi^{z}_{\;j}\Phi^{j}_{\;k}\frac{\hat p^k}{m_a} \label{eqn:relativecoord},
\end{align}
where we have dropped terms past second order in the gravitational potential, assuming $\xi^2 \tau^4\ll 1$. At the end of the protocol, for a fixed bath configuration, we get a ground state population
\begin{align}\label{eqn:curvaturepop}
    \rho_{\downarrow \downarrow}&=\frac{1}{2}+\frac{1}{2}\;\text{Re}
    \;\langle\psi_\downarrow(0)| \exp\{i(\hat \Theta+\theta_0)\} |\psi_\downarrow(0)\rangle
\end{align}
\begin{align}
    \hat \Theta\equiv \left(\frac{G\,\Phi^z_{\;z}\,\tau^2}{2}+\frac{G^2\,\Phi^z_{\;j}\Phi^j_{\;z}\,
    \tau^4}{8} \right)\frac{\hbar  k^2 \tau}{m_a}+k\big(&  G\,\Phi^z_{\;i}  \tau^2+ \frac{7}{12 }G^2\,\Phi^z_{\;j}\Phi^j_{\;i}\tau^4\big)\,\hat r^i\nonumber\\
    &+ \frac{k\tau}{m_a}(G\,\Phi^z_{\;i} \tau^2+\frac{1}{4}G^2\,\Phi^z_{\;j}\Phi^j_{\;i}\tau^4)\,\hat p^i+ \mathcal{O}(\xi^3 \tau^6),
\end{align}
\noindent where again repeated indices are summed over $\{x,y,z\}$. The operator $e^{i\hat \Theta}$ is a phase space displacement, and depends on the bath configuration. 
\newline \indent 
We take for our initial real space wavefunction a Gaussian wavepacket with isotropic width $\sigma$, centered in phase space at position $\mathbf{z_0}=z_0\, \mathbf{e_3}$ and at zero momentum. Denoting our initial state $|\mathbf{z_0}, 0\rangle $, we find
\be
\langle \mathbf{z_0}, 0| e^{ i \hat \Theta } |\mathbf{z_0}, 0\rangle \approx \exp\left\{ik\Big(z_0+\frac{\hbar k \tau}{2m_a} \Big) \tau^2G\,\Phi^z_{\;z}+ \left(i\frac{7 k z_0 }{12 }+i\frac{\hbar k^2\tau}{8m_a} - \frac{ k^2\sigma^2}{2}-\frac{\hbar^2 k^2\tau^2}{8m_a^2\sigma^2}\right) \tau^4 G^2\,\Phi^z_{\;j}\Phi^j_{\;z}\right\},\label{eqn:preBathAve}
\ee
again keeping only terms up to second order in the interaction, which requires the additional assumption $ k^2 \sigma^2\, \xi^2 \tau^4 \ll 1 $. The dominant term in this expression is
\begin{equation}
 \mathscr{D} \equiv \exp\{ik d \tau^2G\,\Phi^z_{\;z}\},
\end{equation} 
a phase shift linear in $\Phi^z_z$ and in the quantity $d\equiv z_0+\hbar k\tau/2m_a$, the distance between the laser and the center of atomic motion. We can think of this as setting an effective dipole moment for the interferometer's response to gravity. Following this there is another phase shift quadratic in $\Phi$, and two inherent decoherence terms set by the nonzero width in phase space of the atomic wavepacket. The sum of these last two terms has a minimum size set by the standard quantum limit at wavepacket width $\sigma = \sqrt{\hbar\tau/2m_a}$.
\newline\indent
At this order, since we are working in the regime where the second derivatives of the gravitational field $\Phi^{z}_{\;j}$ are Gaussian distributed according to Eqn. \ref{eqn:curvVariance}, we can easily trace out the bath as well. We get the following result for the ground state population after averaging over bath configurations, accurate to order $\xi^2\tau^4$:
\begin{align}\label{eqn:curvResultUgly}
    \rho_{\downarrow\downarrow} = \frac{1}{2}+\frac{1}{2}  (\frac{\exp\left\{-\frac{8\pi}{15}k^2 d^2 \; \xi^2\tau^4\right\}e^{i\theta_0}/2}{\sqrt{\prod\limits_{j\in\{x,y,z\}} \Big(1 + \frac{4\pi}{3}\frac{3+\delta_{z\, j}}{5}(k^2 \sigma^2+\frac{\hbar^2 k^2 \tau^2}{4m_a^2\sigma^2}-i\frac{7 k z_0 }{6 }-i\frac{\hbar k^2\tau}{4m_a})\xi^2\tau^4\Big)}} + \text{c.c.}),
\end{align}
where c.c. stands for complex conjugate and we fix the branch of the square roots in Eqn. \ref{eqn:curvResultUgly} by demanding that they continuously go to unity as $\tau\to 0$. Since the real part of the argument is always positive, this eliminates any ambiguity. Tracing the different terms in Eqn. \ref{eqn:preBathAve} through the general formula for a Gaussian function integrated against a Gaussian probability distribution, we see that the linear-in-$\Phi$ phase shift and the inherent decoherence terms lead to a reduction in overall contrast after averaging over bath configurations, while the phase shift quadratic in $\Phi$ actually leads, at lowest order, to an average phase shift in the interferometer, i.e.
\begin{align}\label{eqn:curvResult}
    \rho_{\downarrow\downarrow} \approx \frac{1}{2}+\frac{1}{2}\frac{\exp\left\{-\frac{8\pi}{15}k^2 d^2 \; \xi^2\tau^4\right\}}{\sqrt{\prod\limits_{j\in\{x,y,z\}} \Big(1 + \frac{4\pi}{3}\frac{3+\delta_{z\, j}}{5}(k^2 \sigma^2+\frac{\hbar^2 k^2 \tau^2}{4m_a^2\sigma^2})\xi^2\tau^4\Big)}}  \cos\left(\theta_0+ \frac{4\pi}{3}k(\frac{7}{6}z_0+\frac{\hbar k\tau}{4m_a})\xi^2\tau^4\right).
\end{align}
\indent
Usually $z_0\gg \sigma$, so the factor of $\exp\{-\frac{8\pi}{15}\,k^2 d^2 \, \xi^2\tau^4\}$ provides most of the decoherence. From the dependence on the effective dipole moment we see explicitly how increasing the spatial extent of the experiment increases its sensitivity to effects from finite spacetime curvature, as is familiar from discussions of Einstein's elevator. We also see that the atom-laser separation and the maximum atom path separation both play a role in determining the sensitivity to gravitational noise. Note that the quantities $k$ and $d$, i.e. the inverse wavelength of the wave being interfered and the spatial extent of the experiment, have direct analogues in an interferometer of any kind, while $\xi^2$ of course depends only on the bath. We therefore expect that something like this dominant decoherence term will show up even in interferometers that use a different kind of wave, such as optical interferometers.
\newline \indent
We now plug in typical parameters to get some quantitative understanding of the decoherence behavior. Looking at the dominant contribution to decoherence in Eqn. \ref{eqn:curvResult} mentioned above, we see that $t_{decoherence} \equiv \big(k^2 d^2 \xi^2\big)^{-1/4}$ sets a characteristic decoherence timescale. To get an order-of-magnitude estimate of this timescale, we take $r_{\rm min}$ to be comparable to $d$. Let us consider dark matter providing the background noise. Since the local mass density of dark matter is fixed by observation (at $m_b n_0\approx 5\times10^{-25}$ g/cm$^3$) \cite{Read_2014}, the decoherence timescale decreases as we consider increasingly massive dark matter candidates. Taking an optical wavelength of 780 nm and a large dark matter mass of the Planck mass, $m_{\rm pl}$, to get an optimistic estimate of the decoherence timescale, we get
\begin{equation}
    t_{decoherence} \approx  10 \text{ years}\, \Big(\frac{m_{\rm pl}}{m_b}\Big)^{\frac{1}{4}}\Big(\frac{5\times10^{-25}\text{ g/cm}^3}{m_b n_0}\Big)^{\frac{1}{4}}\Big(\frac{d}{1 \text{ m}}\Big)^{\frac{1}{4}}
\end{equation}
which is clearly unobservable in the near term. Note, too, that in an experiment this long, several of the assumptions that went into the calculation would be violated. Note also that the $\tau^4$ behavior of the exponent in Eqn. \ref{eqn:curvResult} makes it very difficult to see decoherence for experiments with duration much shorter than the characteristic decoherence time.

\subsection{Bias in Gravitational Field}

Above, we analyzed the case where the background gas is distributed isotropically, so that on average the gravitational field in the region of the experiment is zero and in particular, $\langle \Phi^i\rangle = \langle  \Phi^i_{\;j}\rangle = 0$. For completeness, we now analyze the effect of adding a static asymmetry to the gravitational field in the vicinity of the experiment. Though we will put forward a concrete source model later, at the level of the potential second derivatives $\Phi^i_{\;j}$ we assume that the variables are still Gaussian distributed with the variances calculated earlier and proceed to show what happens if they are allowed to have nonzero means. Note that all effects of the single derivatives $ \Phi^i$ disappear even with nonzero mean values, as these terms lead to global, and therefore unobservable, acceleration. Assuming the expectation values are about the same size as the fluctuations, so that we still require calculations to be accurate to order $G^2$,  Eqn. \ref{eqn:curvResultUgly} becomes
\begin{align}
    \rho_{\downarrow\downarrow} = \frac{1}{2}+\frac{\exp\{-\frac{A^2}{2} \text{Var}(\Phi^z_{\;z})-B(\langle \Phi^z_{\;x}\rangle^2+\langle \Phi^z_{\;y}\rangle^2+\langle \Phi^z_{\;z}\rangle^2)\}}{2}\times\nonumber\\
    &\hspace{-2.5in}\left(\frac{\exp\{i\left(\theta_0+A\langle\Phi^z_{\;z}\rangle+C(\langle \Phi^z_{\;x}\rangle^2+\langle \Phi^z_{\;y}\rangle^2+\langle \Phi^z_{\;z}\rangle^2)\right)\}/2}{\sqrt{\prod\limits_{j\in\{x,y,z\}} \Big(1 + \frac{8\pi}{3}\frac{3+\delta_{z\, j}}{5}(B-iC)\xi^2\tau^4\Big)}} + \text{c.c.}\right),
\end{align}
with
\begin{align}
    A &\equiv k d \tau^2 G,\\
    B &\equiv (\frac{k^2 \sigma^2}{2}+\frac{\hbar k^2\tau^2}{8m_a^2 \sigma^2})\tau^4 G^2,\\
    C &\equiv (\frac{7k z_0}{12}+\frac{\hbar k^2 \tau}{8m_a})\tau^4 G^2,
\end{align}
where Var means the variance, i.e. $\text{Var}(\Phi) = \langle\Phi^2\rangle -  \langle\Phi\rangle^2$. We consider a built-in angular asymmetry in the neighborhood of the experiment, of characteristic size $R'>r_{\rm min}$, which we implement for ease of calculation as a hard cutoff on the distribution of particles. That is, in addition to the stochastic background of the previous section, we add a fixed bath mass distribution 
\begin{equation}
    \rho_{\textrm{asym}}(\mathbf{r}) = \begin{cases}
    0, & r<r_{\rm min}\\
    \sqrt{\frac{5}{16\pi}} \,m_b n_{\textrm{asym}} Y_{2,0}(\theta,\phi), & r_{\rm min}<r<R'\\
    0, & R'<r
    \end{cases}
\end{equation}
where $n_{\textrm{asym}}$ is some number density setting the amplitude of the built-in asymmetry, and $Y_{2,0}(\theta, \phi) = \sqrt{5/16\pi} (-1+3 \cos^2\theta)$ is the spherical harmonic of degree $2$ and order $0$. This gives rise to a nonzero expectation value for $\Phi^z_{\;z}$, the second $z$-derivative of the gravitational potential, such that
\begin{equation}
    \langle\Phi_{\;z}^z\rangle  = \frac{5}{8} m_b n_{\textrm{asym}} \int_{r_{\rm min}}^{R'}   \frac{dr}{r} \int d\!\cos \,\theta  (-1+3 \cos^2 \theta)^2 = m_b n_{\textrm{asym}} \ln\frac{R'}{r_{\rm min}}.
\end{equation}
Plugging in $k = 2\pi/270 \text{ nm}$ and again taking $r_{\rm min} \approx d$, the condition $A\langle \Phi^z_{\;z} \rangle  \approx 1$ gives the timescale $t_{\textrm{phase}}$ on which a significant phase shift accumulates:
\begin{equation}
    t_{\textrm{phase}} \approx 5,000 \text{ years}\,\Big(\frac{5\times10^{-25}\text{ g/cm}^3}{m_b n_{\textrm{asym}}}\Big)^{\frac{1}{2}}\left( \frac{1\text{ m}}{d}\right)^{\frac{1}{2}} \left(\frac{1}{\ln \frac{R'}{10^{20} \text{ m}}+\ln \frac{1\text{ m}}{d} }\right)^{\frac{1}{2}}.
\end{equation}
In principle this timescale becomes arbitrarily small as the spatial extent of the asymmetry $R'$ increases, but in practice the logarithmic behavior means that at the local dark matter mass density, even a galaxy-sized asymmetry of $R' = 10^{20}$ m is unobservable. We also note that in practice the gravitational field from nearby stars and planets will contribute to the expectation values $\langle \Phi^i_{\;j}\rangle$.

\section{Relationship to Collisional Brownian Motion}\label{sec:CD}

As discussed in the introduction, dephasing from random baths of background particle collisions with a superposed mass is usually treated through a Brownian motion approach \cite{Gallis1990,Hornberger2003a,Adler2006}. There, the assumption is usually made that the superposed particle is sufficiently heavy that we can ignore changes to its kinetic energy from the scattering, i.e. in the limit $m_a\ll m_b$. The decoherence rate $\Gamma$ for an object held in a superposition of two locations with separation $\Delta x$ is then given by the following integral over $q$, the incoming bath particle momentum: 
\begin{equation}
\Gamma = n \int_0^\infty dq f(q, T) \frac{q}{m_b} \sigma(q),
\end{equation}
in the limit that distance between the two locations is much larger than the typical bath particle de Broglie wavelength, $\Delta x\sqrt{m_b k_B T} /\hbar \gg 1$. Here $n$ is the bath particle number density, $f(q,T)$ is the Boltzmann distribution at temperature $T$, and $\sigma(q)$ is the total scattering cross-section.
\newline \indent
With an unscreened $1/r$ interaction, the above poses an immediate problem: the total cross section is infinite. This can be traced back to the long-range potential: the potential does not turn off sufficiently fast even as the bath particle moves arbitrarily far away, and the particle continues to scatter at infinitely late times \cite{Dollard1964}. 
\newline\indent
The divergences associated with the $1/r$ potential occur even at the semiclassical level. For instance, a bath wavepacket far from the scattering center will continue, even neglecting any spread in the wavepacket about its mean location, to accumulate an overall phase
\begin{align}\label{eqn:infinitePhase}
    \phi(t) =  \int^t dt' \frac{V_0}{|\mathbf{r} + \mathbf{v} t'|}
\end{align}
which grows without bound as $t \to \infty$. Above, the wavepacket is moving in a potential centered at the origin, $\mathbf{r}$ is the initial location of the wavepacket, $\mathbf{v} $ is its asympotic velocity, $V_0$ is a coupling constant, and we are focusing on the behavior when $|\mathbf{v} t |\gg |\mathbf {r}|$.
\newline \indent
However, in seeking to model the decohering effect of a distant bath particle upon a spatial superposition, there is a related quantity which remains finite, i.e. the difference of phases. A bath particle produces the relative phase shift between the two atom paths of the form
\begin{equation}
\Delta \phi \approx V_0 \int_{0}^{t} dt' \left( \frac{1}{|\mathbf{r}_1(t') - \mathbf{r} - \mathbf{v} t'|} - \frac{1}{|\mathbf{r}_2(t') - \mathbf{r} - \mathbf{v} t'|} \right).
\end{equation}
Here, $\mathbf{r}_{1,2}$ represent the two locations of the atom along the superposition path, $V_0$ is a coupling constant, and $\mathbf{r}, \mathbf{v}$ are, as in Eqn. \ref{eqn:infinitePhase}, the initial position and velocity of the bath particle, respectively. At sufficiently long times $t \to \infty$ and assuming bounded $r_1, r_2$, the late-time behavior of this integral is
\begin{equation} \label{eqn:phaseDifference}
\Delta \phi =  V_0 \int^{t \to \infty} \frac{dt'}{|\mathbf{v}|^2 t'^2} < \infty
\end{equation}
because the lowest-order terms, which would have diverged logarithmically, cancel. Thus there is no unregulated infrared divergence. This suggests that the resolution to the failure of naive collisional decoherence lies in carefully working in terms of regulated quantities, i.e. quantities defined in terms of the difference of the evolution between two paths.
\newline \indent
To see how this shows up in the calculations already carried out, consider the following: if we followed the typical path integral method of calculation, the first thing to consider would be the phase associated with the potential difference between the two paths
\begin{align}
    \Delta \phi &= \int( V (\mathbf{r}_1(t)) -  V(\mathbf{r}_2(t)))dt\\
    &\approx \int dt \mathbf{F}(t)\cdot \Delta \mathbf{r}(t)
\end{align}
where $\mathbf{F} = - \nabla V$ and in the second line we assume that the force in the vicinity of the experiment is approximately constant in space. This of course is not accurate for bath particles close to the experiment, but the infrared divergence discussed above comes from the behavior of the interaction at large distances, so it is sufficient to study this case. Taking, as in the protocol outlined above,
\begin{equation}
    \Delta \mathbf{r} (t)  = 
    \begin{cases}
    kt/m_a \,\mathbf{e_3} & t \in [0,\tau]\\
    k(2\tau-t)/m_a \,\mathbf{e_3} & t \in [\tau,2\tau],
    \end{cases}
\end{equation}
we have
\begin{align}
    \Delta\phi&\approx \int_\tau^{2\tau} dt F_z(t) (k (2\tau-t)/m_a)+ \int_0^\tau dt F_z(t) (k t/m_a) \\
    &= k\left(\int_0^{2\tau}\! \int_0^{t'} dt' dt'' \,\frac{F_z(t'')}{m_a} -2 \int_0^{\tau}\! \int_0^{t'} dt' dt'' \,\frac{F_z(t'')}{m_a}\right)\\
    &\approx k \left(z(2\tau)-2 \,z(\tau)\right)
\end{align}
which is of course the semiclassical version of Eqn. \ref{eqn:intOverlap} neglecting laser motion. We therefore stress that we evade divergences associated with the $1/r$ potential not simply because we model the bath classically (recall that divergence occurs even at the semiclassical level), but rather because we rigorously track \textit{relative} evolution of the atomic superposition, in a way that will remain valid even beyond this lowest term in the semiclassical expansion of bath particle behavior. Our approach also goes beyond the infinite experimental mass limit of standard collisional decoherence, i.e. we account for how interaction with the bath warps the two paths of the interferometer. Finally, note that the finite result in Eqn. \ref{eqn:phaseDifference} suggests there is a scattering treatment of the problem one could consider applying in the case where the atom is held at two fixed locations for a long time, as in \cite{doi:10.1126/science.aay6428}.

\section{Collision Cone Sector}\label{sec:Near}
We now consider the effect of bath particles in the collision cone on atom interferometry signals. Recall that the collision cone consists of those background particles which at some point during the atom freefall come within the cutoff radius $r_{\rm min}$. These particles come close to the setup in the sense that they source a potential which may vary significantly across the extent of the experiment. While closed-form expressions for the scattering states of single particles in the Newton potential exist \cite{Dollard1964,Herbst1974}, we use an approximation motivated by dark matter detection and focus on the impulse delivered by collision cone particles on the components of our interferometer.
\newline \indent 
Our approximation neglects self-consistent corrections that lead to high angle scattering. Specifically, in 1/r scattering, high angle scattering events correspond to substantial back-action on the incoming particle. We neglect the change in the motion of the bath particle to all orders, and keep the only the lowest order correction to changes in the motion of the interferometer. We also neglect quantum back-action, i.e. entanglement between the bath particle state and the atom state.
\newline \indent 
For simplicity we consider the effect of a single particle in the collision cone. In order to refer to the center of atomic motion more conveniently, we define 
\begin{align}
 \mathbf d \equiv \mathbf z_0 + \frac{\hbar \mathbf k \tau}{2 m_a}.
\end{align}
We use the impulsive limit to write down simple expressions for the force on the atom and the laser. Given the bath particle initial conditions we first calculate the times at which it comes nearest to the initial laser location $\mathbf 0$ and the average atom location $\mathbf d$. This done, rather than write out the full expression for how the force changes continuously in time, we model it as instantaneous kicks delivered to the experimental components at the times of closest approach. With this model the force equations are
\begin{align}
    &\dot{P}_{z,\,laser} = M_{\rm laser} \,\delta v_z(\mathbf 0) \,\delta\big(t-t_{\text{kick}}(\mathbf 0)\big)\\
    &\dot{\hat p}_z =  m_a \,\left[\delta v_z (\mathbf d)+ \nabla \big( \delta v_z(\mathbf d)\big) \cdot (\hat{ \mathbf{r}}-\mathbf d) + \mathcal{O}((\hat{\mathbf r}-\mathbf d)^2) \right] \,\delta\big(t-t_{\text{kick}}(\mathbf d)\big), \label{eqn:atomKick}
\end{align}
where  $t_{kick}(\mathbf r)$ is the time at which a bath particle with position $\mathbf{r_b}$ and velocity $\mathbf{v_b}$ at $t=0$ comes closest to the point $\mathbf r$
\begin{align}
    t_{\text{kick}}(\mathbf r)&\equiv -\frac{(\mathbf{r_b}-\mathbf r)\cdot\mathbf{v_b}}{v_b^2}
\end{align}
and the velocity kick function is defined by integrating the gravitational acceleration felt at $\mathbf r$ over time
\begin{align}\label{eqn:velocityKick}
    \delta v_z(\mathbf r) \equiv \int_{-\infty}^\infty dt \frac{G m_b (\mathbf{r_b}+\mathbf{v_b}t - \mathbf r) }{|\mathbf{r_b}+\mathbf{v_b}t - \mathbf r|^3} = \frac{2 G m_b}{v_b}\frac{\Big((\mathbf{r_b}-\mathbf{r})+\mathbf{v_b}t_{\text{kick}}(\mathbf r)\Big)\cdot \mathbf{e_3}}{\Big((\mathbf{r_b}-\mathbf{r})^2-v_b^2 t^2_{\text{kick}}(\mathbf r)\Big)}.
\end{align}
Note the scaling $\delta v_z \propto G m_b/v_b b$, where $b$ is the impact parameter of the bath particle trajectory with respect to $\mathbf{r}$. The velocity kick function has gradient
\begin{align}
    \nabla(\delta v_z(\mathbf r)) &= \frac{4 G m_b}{v_b}\frac{\Big((\mathbf{r_b}-\mathbf{r})+\mathbf{v_b}t_{\text{kick}}(\mathbf r)\Big)\cdot \mathbf{e_3}}{\Big((\mathbf{r_b}-\mathbf{r})^2-v_b^2 t^2_{\text{kick}}(\mathbf r)\Big)^2}\Big((\mathbf{r_b}-\mathbf{r})+\mathbf{v_b}t_{\text{kick}}(\mathbf r)\Big) \nonumber \\[.2in] 
    & \hspace{1.5in}-\frac{2 G m_b}{v_b}\frac{\mathbf{e_3}-\mathbf{v_b} (\mathbf{v_b}\cdot \mathbf{e_3})/v_b^2}{\Big((\mathbf{r_b}-\mathbf{r})^2-v_b^2 t^2_{\text{kick}}(\mathbf r)\Big)}.
\end{align}
Because we assume that the bath particle still remains much farther than $k\tau/m_a$ from the atom at all times, the velocity kick felt by the atom at any point in its trajectory is well approximated by a linear correction to the velocity kick felt at $\mathbf d$. Going forward we will drop the terms beyond linear order in $\hat{\mathbf r}-\mathbf d$. 
\newline \indent 
We can write down the evolution of the laser and atom positions
\begin{equation}
    Z(t) = \Theta(t-t_{\text{kick}}(\mathbf 0)) \,(t-t_{\text{kick}}(\mathbf 0)) \,\delta v_z(\mathbf 0)
\end{equation}
\begin{equation}
\hat z(t) = \hat z + \frac{\hat p_z t}{m_a} +\Theta(t-t_{\text{kick}}(\mathbf d))\,(t-t_{\text{kick}}(\mathbf d))\,\left[\delta v_z(\mathbf d)+\nabla( \delta v_z(\mathbf d))\cdot(\mathbf{\hat r}+\frac{\mathbf{\hat p}\, t_{\text{kick}}(\mathbf d)}{m_a}-\mathbf d)\right]
\end{equation}
where $\Theta(t)$ is the unit step function which results from integrating the delta function in time. The final overlap factor, which we use to determine the final ground state population according to Eqn. \ref{eqn:Population}, is
\begin{align}
\langle \mathbf{z_0}, 0| (U^b)^\dag U^t|\mathbf{z_0}, 0\rangle =& e^{i \theta_0} \exp\left\{ik\,\Big(\delta v_{z}(\mathbf d)\,(\tau-|\tau - t_{\text{kick}}(\mathbf d)|)- \delta v_{z}(\mathbf 0)\,(\tau-|\tau - t_{\text{kick}}(\mathbf 0)|)\Big)\right\}\nonumber\\
& \quad \times \exp\left\{i \frac{\hbar k^2 }{2 m_a}\,\partial_z\big(\delta v_z(\mathbf d)\big)\,(t_{\text{kick}}(\mathbf d)-\tau)\,(\tau-|\tau-t_{\text{kick}}(\mathbf d)|)\right\}\nonumber\\
&\quad\times\exp\left\{-\nabla\big(\delta v_z(\mathbf d)\big)^2(\tau-|\tau-t_{\text{kick}}(\mathbf d)|)^2\left(\frac{k^2 \sigma^2}{2}+\frac{\hbar^2 k^2 t_{\text{kick}}^2(\mathbf d)}{8 m_a^2 \sigma^2 }\right)\right\}. \label{eqn:collConeSig}
\end{align}
Let us examine the features of this expression. The first line gives a simple phase shift caused by relative motion of the atom and the laser. Note that a velocity kick to the laser system is just as easy to read out in the signal as a kick to the atom, which emphasizes that both components of the experiment contribute to its function as a sensor. The second term is a more complicated phase shift depending on the differential velocity kick to the two atom paths. The third term gives an overall reduction in contrast since the differential velocity kicks to the two atom paths cause the different paths to end at slightly different positions and momenta, inhibiting their ability to interfere. 
\newline \indent 
As the time of the velocity kicks approaches the beginning or end of the experiment $t_{\text{kick}}(\mathbf 0),t_{\text{kick}}(\mathbf d) \to 0, 2\tau$, their effect disappears. To explain, at the beginning and end of the experiment the separation between atom paths is negligible and so the differential velocity kick is also negligible, while the relative motion of the laser and atom leads either to identical and therefore unobservable phase shifts on both arms ($t_{kick}\to 0$) or to negligible phase shifts ($t_{kick} \to 2\tau$). Notice that if the average kick to the atom equals the laser kick $\delta v_z(\mathbf d) = \delta v_z(\mathbf 0)$, and if these occur at the same time $t_{kick}(\mathbf 0) = t_{kick}(\mathbf d)$, then the phase shift resulting from relative motion between the atom and laser again disappears, but effects related to different velocity kicks on the two arms (set by $\nabla (\delta v_z)$, the velocity kick gradient) persist.
\newline \indent
We now comment on prospects for observing this gravitational noise. The dominant contribution to Eqn. \ref{eqn:collConeSig} is the phase shift in the first line whose characteristic size scales with experimental parameters like
\begin{equation}
    \Delta \theta \approx k\, \delta v_z \tau.
\end{equation}
Obviously, a measurement on a single atom is subject to shot noise and cannot give meaningful information on a small phase shift if the phase shift is not constant over many runs of the experiment. With $N$ atoms in a single cloud, however, one reduces the shot noise by a factor of $\sqrt{N}$. The velocity sensitivity of the interferometer, then, scales like $(Q k\tau\sqrt{N})^{-1}$, where $\hbar k$ is the single photon momentum and $Q$ is an integer to include the increased momentum kick of multiphoton transition pulses  \cite{Kovachy2015,Dickerson2013,Miffre2006,Overstreet2021,Chiow2011}. With reasonable experimental numbers the weakest observable kick, $\delta v_{z,\, \rm{min}}$, is roughly
\begin{equation}\label{eqn:sensitivity}
    \delta v_{z,\, \rm{min}} \approx 10^{-12}\,\frac{\text{m}}{\text{s}}\, \Big(
    \frac{10^2}{Q}\Big)\Big(
    \frac{2\pi/780 \text{ nm}}{k}\Big)\Big(\frac{1 \text{ s}}{\tau}\Big)\Big(\frac{10^6}{N}\Big)^{1/2},
\end{equation}
that is, this kick produces a phase shift on the order of 1 radian. For comparison, a heavy dark matter particle (with a mass of $100$ $m_{\rm pl}$) passing within tens of meters of the atom produces a velocity kick on the order of $10^{-22}$ m/s. From the scaling of the velocity kick given just after Eqn. \ref{eqn:velocityKick}, $\delta v_z \propto b^{-1}$, we see that in order to produce a measurable phase shift the same dark matter particle would need to pass within nanometers of the atom. An event of this type has negligible event rate and would of course require modelling the potential near the center of atomic motion beyond linear order in $\hat{\mathbf r}-\mathbf d$.

\section{Tuning the Sensor: the Choice of Atom-Laser Distance}\label{sec:Tuning}
We now comment on the role of the tunable parameter $d \equiv z_0 +\hbar k\tau/2m_a$, the distance between the laser and the center of atomic motion. In practice, the easiest way to adjust this is through $z_0$, the initial atom-laser distance. Recognizing, as we have emphasized, that the laser system is part of the sensor, we now ask, what goes into optimizing the atom-laser distance for sensitivity to the gravitational background?

\begin{figure}
    \centering
    \subfloat[]{
        \includegraphics[width=.48\linewidth]{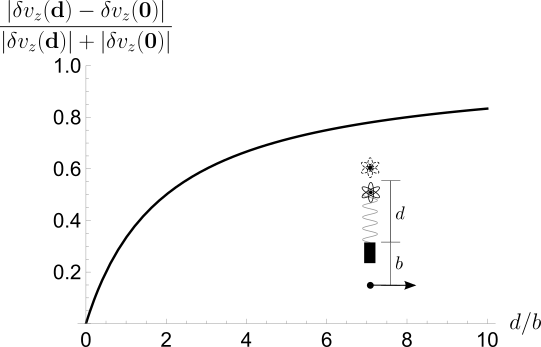}\label{fig:Fig3a}
    }
    \subfloat[]{
        \includegraphics[width=.48\linewidth]{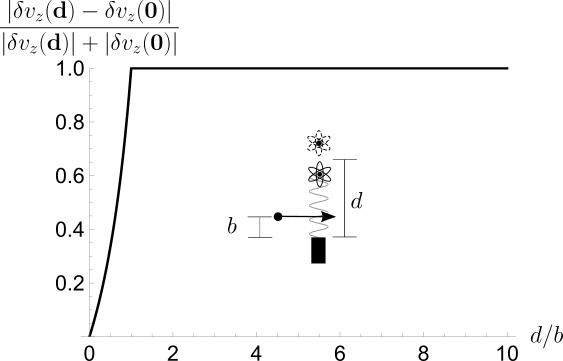}\label{fig:Fig3b}
    }
    \caption{Plot of the magnitude of the dominant phase shift in Eqn. \ref{eqn:collConeSig} divided by the sum of the constituent phase shift magnitudes. For $d\ll b$ there is significant phase shift cancellation due to similar common motion of the laser and the atom. a) A bath particle passes below the laser. The size of the net phase shift is about 80\% of the sum of the separate phase shift magnitudes for $d = 10b$, the ratio asymptotes to 1 in the limit $d/b\to \infty$. Increasing $d$ far beyond $b$ gives diminishing returns. b) A bath particle passes above the laser. The behavior resembles the previous case for $d/b\ll1$, but as soon as $d > b$, the momentum kicks on the atom and laser are in opposite directions, so there is no common motion cancellation to the dominant phase shift.}
\end{figure}

Recall that we found it useful to distinguish between two sectors of bath particle phase space, the distant sector and the collision cone. We will take these sectors in turn. In fact, however, the most significant change to effects from the distant sector when we change $d$ is the redefinition of the distant sector itself. Physically, the atom-laser distance is important if one wants to see gravitational effects because it sets the cutoff beyond which effects of the bath have significant common motion cancellations. The notion of a distant sector therefore gets its meaning from the length scale $d$.
\newline\indent
Mathematically, the dependence on $d$ appears in the formula for decoherence resulting from distant sector particles, Eqn. \ref{eqn:curvResult}, both in an explicit factor of $d = z_0+\hbar k\tau/2m_a$ and implicitly in the characteristic size of gravitational field fluctuations $\xi^2 \propto 1/r_{\rm min}^3$, through the condition $r_{\rm min} \gg d$. Since $r_{\rm min}$ must scale up with $d$ for the cutoff radius to contain the entire apparatus, the net effect is that contribution of distant sector particles to decoherence is diminished as the spatial extent of the experiment is increased. This makes intuitive sense. As we increase $d$, more and more particles we previously labeled as ``distant'' we now label as being in the collision cone. The decoherence from these particles still matters at large $d$, but is now attributed to the collision cone contribution.
\newline\indent
The sensitivity of the interferometer to collision cone particles as we increase $d$ is more interesting. The dominant contribution in the formula for the effect of collision cone particles, Eqn. \ref{eqn:collConeSig}, is a sum of the phase shift from the impulse on the laser and the phase shift from the mean impulse on the atom. Consider for simplicity a bath particle with velocity perpendicular to the $z$ direction which passes through the point $(0,0,-b)$ at time $\tau$. Note that we focus on $-b<0$ so that in the large $d$ limit the dominant effect we need to analyze is the velocity kick to the laser. This is simpler to treat than the effect on the atomic superposition, but, as pointed out in the previous section, is still readily observable. Take the impact parameter $b$ to be fixed, and consider changing the sensitivity of the signal to this bath particle by varying $d$. As shown in Fig. \ref{fig:Fig3a}, in the regime $d \ll b$, the magnitude of the net phase shift is much smaller than the sum of separate magnitudes of the laser shift and average atom shift (there is significant cancellation between the laser shift and atom shift), but as $d/b \to \infty$ the ratio of the two asymptotes to 1 (one of the shifts is much larger than the other). Over 80 \% of the maximum recoverable phase shift is achieved when $d/b = 10$.
\newline\indent 
Very roughly, then, we can say that the signal is not significantly reduced by common motion cancellations for bath particles with impact parameter $\lesssim d$. A larger choice of $d$ leads to greater sensitivity to particles with impact parameters on the order of the experiment size. However, since a larger impact parameter leads to weaker velocity kicks to the apparatus, at some point increasing $d$ further "brings online" particles whose effect is too weak to be read out. Increasing $d$ far beyond this length scale provides no advantage to the experiment. With the simplifications we have made in this section, the idea of a minimum velocity kick sensitivity from Eqn. \ref{eqn:sensitivity} naturally leads to a maximum impact parameter $b_{max}$ for bath particles in the collision cone such that their effect can be read out, i.e. the condition on $d$ such that the only momentum kicks which suffer common motion cancellations are those already too small to be read out is
\begin{equation}
    d \gg b_{max} \approx 1\text{ nm}\;\Big(\frac{m_b}{m_{\rm{pl}}}\Big)\Big(\frac{10^{-14} \text{ m/s}}{\delta v_{z,\, \rm{min}}}\Big).
\end{equation}
There is no advantage to increasing $d$ once it is already much larger than $b_{max}$, which we see above is automatically fulfilled in any realistic experiment. As a result, there is little practical significance in the choice of $d$.

\section{Cosmic Rays}\label{sec:Cosmic}

Finally, we consider the atom interferometer as a velocity kick detector for the case of bath particles coupled through electromagnetic forces. Note in this case, we do not need to track the evolution of the laser, which we take to be charge neutral and very massive. Consider a cosmic ray particle with charge $q$ at location $\mathbf{r}_c(t)$. We neglect the many-electron internal structure of the atom, considering only a single valence electron and a nucleus of effective charge $+e$. Again, $\hat{\mathbf{r}}$ is the operator for the atomic center of mass. We also now need to consider $\hat{\mathbf{r}}'$, the relative coordinate operator, which gives the displacement of the electron from the nucleus. In the nonrelativistic limit, the interaction Hamiltonian of the atom with the cosmic ray particle is
\begin{align}
    \hat H'  &= \frac{qe}{4 \pi \epsilon_0|\hat{\mathbf{r}}-\mathbf{r}_c|}-\frac{qe}{4 \pi \epsilon_0 |\hat{\mathbf{r}}+\hat{\mathbf{r}}' -\mathbf{r}_c|}\\
    &\approx - \frac{qe}{4 \pi \epsilon_0|\hat{\mathbf{r}}-\mathbf{r}_c|^3} (\mathbf{r}_c-\hat{\mathbf{r}})\cdot \hat{\mathbf{r}}'\\
    &=-\hat{\mathbf{d}}\cdot \mathbf{E}(\hat{\mathbf{r}}), \qquad \hat{\mathbf{d}}\equiv  - e \hat{\mathbf{r}}', \; \mathbf{E}(\hat{\mathbf{r}})\equiv - \frac{q}{4 \pi \epsilon_0|\hat{\mathbf{r}}-\mathbf{r}_c|^3} (\mathbf{r}_c-\hat{\mathbf{r}}),
\end{align}
where to get the second line we assume that the distance of the cosmic ray from the atom is always much larger than the atom size. We then get a dipolar interaction of the atom with the electric field produced by the cosmic ray.
\newline \indent
We assume the cosmic ray passes far enough from atom that the electric field near the atom varies slowly in time and therefore we use the DC Stark effect to get in perturbation theory
\begin{equation}
    \hat H' \approx -\frac{\alpha_a}{2} |E(\hat{\mathbf{r}})|^2 = - \frac{\alpha_a}{2}\frac{q^2}{(4 \pi \epsilon_0)^2|\hat{\mathbf{r}}-\mathbf{r}_c|^4},
\end{equation}
where $\alpha_a$ is the ground state polarizability of the atom. Note that now the energy depends only on the location of the atom center of mass.
\newline \indent
The leading order effect for a cosmic ray particle much farther than the separation between atom paths is just the average velocity kick to the atom. This still produces a measurable phase shift because the atom moves relative to the laser. With impact parameter $ \mathbf{b} \propto \mathbf{e_3}$ and a straight-line charged particle trajectory we get
\begin{align}
    m_{a} \delta v_z &\approx  \alpha_a (\frac{q}{4\pi\epsilon_0})^2 \int_{-\infty}^\infty dt \frac{1}{(\sqrt{b^2+v^2 t^2})^5}\\
    \delta v_z &\approx \frac{\alpha_a}{m_a} (\frac{q}{4\pi\epsilon_0})^2 \frac{1}{ b^4\,v}.
\end{align}
To get an order-of-magnitude estimate, we plug in $c$ for the cosmic ray particle velocity, and take the atom to be rubidium, which has $\alpha_{\text{Rb}} \approx 4\pi \epsilon_0 \times 50$   \AA\hspace{.5mm}$\!^3$ \cite{PhysRevA.92.052513}. This gives
\begin{align}
    \delta v_z &\approx 10^{-40} \;\frac{\text{m}}{\text{s}} \;\left(\frac{q}{e}\right)^2 \left(\frac{1 \text{ m}}{b}\right)^4.
\end{align}
In section \ref{sec:Near} we say with typical experimental parameters one should be able to get a velocity kick sensitivity of roughly $ 10^{-12}$ m/s. We therefore need a cosmic ray proton to pass within about 0.1 $\mu$m of the atom to be able to read it out. Given the cosmic ray number density of about $10^{-3}/\rm{m}^{3}$ \cite{VALKOVIC20005}, this corresponds to an event rate of about one every 10 years and is therefore negligible as a source of continuous noise. Such an event would also more require careful modelling as it violates our assumption that the passing particle be much further from the experiment than the distance between atom paths.
\newline \indent
We can also consider boosting the interaction with an applied static field $E_{applied}$ on the atom. This enhances the effect of cosmic ray particles with impact parameter parallel to the applied field. The size of the velocity kick now goes like
\begin{equation}
    \delta v_z \approx \frac{\alpha_a}{m_a} \frac{e}{4\pi\epsilon_0 b^2} (E_{applied}+\frac{e}{4\pi\epsilon_0 b^2})\frac{1}{v}.
\end{equation}
Suppose we apply a static field on the order of 1 kV/m to the atom. The contribution from the cosmic ray particle gives a similar electric field at a distance of 1 $\mu$m. From the numbers above, we see that even with the applied field, if the cosmic ray passes at a distance of 10 $\mu$m, the phase shift is already undetectable (though suppressed only by a factor of 100 rather than $10^4$ corresponding to the case without an applied field).

\section{Outlook}\label{sec:Conclusion}
In this paper, we examined an explicit model for gravitational background noise in an atom interferometry experiment. By accounting for the effect of the noise on the control system in addition to the atom, we were able to illustrate how the equivalence principle suppresses the effect of bath particles far from the experiment.  Clearly, it will be difficult to see such gravitational noise in any near term experiment.

\section{Acknowledgments}
We thank Gayathrini Premawardhana for helpful conversations and William McGehee, Eite Tiesinga, and Peter Shawhan for comments on the manuscript. DC is supported by the US Department of Energy under contract DE-AC02-05CH11231 and Quantum Information Science Enabled Discovery (QuantISED) for High Energy Physics grant KA2401032.

\appendix

\section{Time Dependence in the Distant Sector}\label{app:timeDep}
When we include the time dependence of the bath, we get the following evolution of the relative displacement
\begin{align}
    \hat z(t) - Z_{\rm laser}(t) \approx\,&  \hat z + t\,\frac{\hat p_z}{m_a} +   G\,\int_0^t dt_1 \int_0^{t_1 }dt_2\Phi^{z}_{\;j}(t_2)\,(\hat r^j + t_2 \frac{\hat p^j}{m_a} ) \nonumber\\
    &+G^2\,\int^t_0 dt_1\int^{t_1}_0 dt_2\Phi^{z}_{\;j}(t_2)\int^{t_2}_0 dt_3 \int^{t_3}_0 dt_4\,\Phi^{j}_{\;k}(t_4)\,(\hat r^k+\,t_4\,\frac{\hat p^k}{m_a}).
\end{align}

\noindent Using the same initial atomic wavepacket $|\mathbf{z_0}, 0\rangle $ from the main text we find
\begin{align}
\langle \mathbf{z_0}, 0| (U^b)^\dag U^t|\mathbf{z_0}, 0\rangle =& \,e^{i\theta_0}\exp\big\{ik\Big(G \int^{2\tau} dt_1 \int^{t_1}dt_2 \Big[\Phi^z_z(t_2)(z_0+\frac{k t_2}{2 m_a})\nonumber\\
&\hspace{1.6in}+\Phi^z_j(t_2)G\int^{t_2} dt_3 \int^{t_3}dt_4\Phi^j_z(t_4)(z_0+\frac{k t_4}{2 m_a})\Big] \nonumber\\
&\hspace{.7in}- 2 G\int^\tau\! dt_1 \! \int^{t_1} \!dt_2  \Big[\Phi^z_z(t_2)(z_0+\frac{k t_2}{2 m_a})\nonumber\\
&\hspace{1.6in}+\Phi^z_j(t_2)G\int^{t_2} dt_3 \int^{t_3}dt_4\Phi^j_z(t_4)(z_0+\frac{k t_4}{2 m_a})\Big] \Big)\big\} \nonumber \\
 & \times \exp\big\{-\!\Big(G \int^{2\tau} \int^{t_1}dt_2 \Phi^z_j(t_2)\! -\!2 G \int^\tau \int^{t_1}dt_2 \Phi^z_j(t_2) \Big)^2\, \frac{k^2\sigma^2}{2}\nonumber\\
 &\hspace{.7in}-\!\Big(G \int^{2 \tau}\int^{t_1}dt_2 \Phi^z_j(t_2)t_2\!-\!2 G \int^\tau \int^{t_1}dt_2 \Phi^z_j(t_2)t_2 \Big)^2\, \frac{k^2}{8m^2 \sigma^2}\big\},
\end{align}

\begin{align}
\langle \mathbf{z_0},& 0| (U^b)^\dag  U^t|\mathbf{z_0}, 0\rangle \\
&= \,e^{i\theta_0}\exp\big\{ik\Big(G \int^{2\tau} dt_1 \int^{t_1}dt_2 \Big[\Phi^z_z(t_2)(z_0+\frac{k t_2}{2 m_a})+\Phi^z_j(t_2)G\int^{t_2} dt_3 \int^{t_3}dt_4\Phi^j_z(t_4)(z_0+\frac{k t_4}{2 m_a})\Big] \nonumber\\
&\hspace{.7in}- 2 G\int^\tau\! dt_1 \! \int^{t_1} \!dt_2  \Big[\Phi^z_z(t_2)(z_0+\frac{k t_2}{2 m_a})+\Phi^z_j(t_2)G\int^{t_2} dt_3 \int^{t_3}dt_4\Phi^j_z(t_4)(z_0+\frac{k t_4}{2 m_a})\Big] \Big)\big\} \nonumber \\
 & \qquad \times \exp\big\{-\!\Big(G \int^{2\tau} \int^{t_1}dt_2 \Phi^z_j(t_2)\! -\!2 G \int^\tau \int^{t_1}dt_2 \Phi^z_j(t_2) \Big)^2\, \frac{k^2\sigma^2}{2}\nonumber\\
 &\hspace{.7in}-\!\Big(G \int^{2 \tau}\int^{t_1}dt_2 \Phi^z_j(t_2)t_2\!-\!2 G \int^\tau \int^{t_1}dt_2 \Phi^z_j(t_2)t_2 \Big)^2\, \frac{k^2}{8m^2 \sigma^2}\big\},
\end{align}
%
where a square over terms with one free index implies doubling and summing over the free index, i.e. $(\Phi^z_j+\cdots)^2\equiv\Phi^z_j\Phi^j_z+\cdots$. In the main text we analyze the average over bath configurations of the phase shift in the above expression which is linear in $\Phi$, as this term is responsible for the dominant effect in the static case. The dominant term $\mathscr{D}$ from Eqn. \ref{eqn:preBathAve} in the static case was made up of terms that are linear in $\Phi$. In the case of time dependent $\Phi$ this becomes
\begin{align}
\mathscr{D} =\exp\big\{ikG\Big(  \int^{2\tau} \!dt_1 \! \int^{t_1}dt_2 \Phi^z_z(t_2)(z_0+\frac{k t_2}{2 m_a})- 2 \int^\tau \!dt_1 \!  \int^{t_1}dt_2 \Phi^z_z(t_2)(z_0+\frac{k t_2}{2 m_a})\Big)\big\}.\label{eqn:timDepPhase}
\end{align}
Now we assume that we can approximate $\Phi^z_z(t_2)$ with a linear function in time,
\begin{equation}
    \Phi^z_z(t) \approx \Phi^z_z(0) + \partial_t \Phi^z_z(t)\Big|_{t=0}\, t
\end{equation}
where
\begin{equation}
     \partial_t \Phi^z_z(t)\Big|_{t=0} = \,m_b\sum_{n=1}^N \; 3\frac{(r_{b_n}^2-5z_{b_n}^2)\mathbf{r}_{b_n}\cdot\mathbf{v}_{b_n}}{r_{b_n}^7}+6\frac{z_{b_n} v_{z,b_n}}{r_{b_n}^5}.
\end{equation}
This approximation assumes that we are looking at experimental runtimes still much shorter than the autocorrelation time of the bath. In particular, we assume $v_\beta \tau \ll r_{\rm min}$. In this limit, we approximate the integral over the distant sector with an integral over all initial positions outside $r_{\rm min}$ and over all velocities regardless of initial position. As a result the correlation between $\Phi^z_z(0), \partial_t \Phi^z_z(0)$ vanishes by symmetry as $\partial_t \Phi^z_z(0)$ is linear in velocity, and we can treat the two quantities as independent Gaussian random variables, with
\begin{equation}
    \langle (G\,\partial_t \Phi^z_z(0))^2\rangle = \frac{48\pi}{ 5}\xi^2 \frac{v_\beta^2}{r_{min^2}}.
\end{equation}
We can then straightforwardly extend the calculation from the static bath section to get
\begin{equation}
    \langle\mathscr{D}\rangle_{Bath} = \exp\{-\frac{8\pi}{15}k^2(z_0+ \frac{k\tau}{2m_a})^2\xi^2 \tau^4\}\exp\{-\frac{24\pi}{5} k^2(z_0+\frac{7}{12}\frac{k \tau}{m_a})^2\xi^2  \tau^4\frac{ v_\beta^2\tau^2}{ r_{\rm min}^2}\},
\end{equation}
the lowest order correction in bath time dependence.

\bibliography{apssamp}

\end{document}